\documentclass[preprint,10pt,nobibnotes,twocolumn,byrevtex,aps]{revtex4}%
\usepackage{amsfonts}
\usepackage{amsmath}
\usepackage{amssymb}
\usepackage{graphicx}%
\begin{document}
\preprint{AUTP/04-07}
\title{Stationary Metastability in an Exact Non-Mean Field Calculation for a Model
without Long-Range Interactions}
\author{P. D. Gujrati}
\affiliation{Department of Physics, Department of Polymer Science,\ The University of
Akron, OH, USA 44325}
\date{\today}

\begin{abstract}
We introduce the concept of stationary metastable states (SMS's) in the
presence of another more stable state. The stationary nature allows us to
study SMS's by using a restricted partition function formalism as advocated by
Penrose and Lebowitz and requires continuing the free energy. The formalism
ensures that SMS free energy satisfies the requirement of thermodynamic
stability everywhere including $T=0,$ but need not represent a pysically
observable metastable state over the range where the entropy under
continuation becomes negative. We consider a 1-dimensional $m$-component
axis-spin model involving only nearest-neighbor interactions, which is solved
exactly. The high-temperature expansion of the model representys a polymer
problem in which $m$ acts as the activity of a loop formation. We follow
deGennes and trerat $m$ as a real variable. A thermodynamic phase transition
occurs in the model for $m<1.$ The analytic continuation of the
high-temperature disordered phase free energy below the transition represents
the free energy of the metastable state. The calculation shows that the notion
of SMS is not necessaily a consequence of only mean-field analysis or requires
long-range interactions.

\end{abstract}
\maketitle

\section{Introduction}

Supercooled and superheated states are ubiqutious in Nature, even though they
\emph{cannot} be rigorously derived from \emph{equilibrium} statistical
mechanics \cite{Penrose}. Their observation is usually justified by appeal to
the \textquotedblleft van der Waals loop\textquotedblright\ in the celebrated
van der Waals equation for the liquid-gas transition. The existence of the
loop violates the fundamental property that the\ partition function (PF) be
maximized or the free energy be \emph{convex}. Despite this, metastable states
appear not only in many other mean-field theories such as the Bragg-Williams
theory \cite{Bragg}, but can easily be prepared in the\ labratory. There are
usually two different mechnism operative in metastable states. There is
usually a \textquotedblleft fast\textquotedblright\ mechanism (time scale
$\tau_{\text{f}}$) to create a metastable state in the system, followed by a
\textquotedblleft slow\textquotedblright\ mechanism (time scale $\tau
_{\text{s}}$) for nucleation of the stable phase and the eventual decay of the
metastable state. For the metastable state to exists for a while, we need to
require $\tau_{\text{s}}>\tau_{\text{f}}.$ In approximate theories, the
thermodynamic functions for the metastable states are taken as the
extrapolation of the functions from the nearby equilibrium states. However,
metastable states in real systems always have time-dependence associated with
them. Thus, the metastable state represented by extrapolation can only
represent the \emph{stationary} limit ($\tau_{\text{s}}\rightarrow\infty$) of
experimentally observed metastable states; see also \cite{Derrida0}. However,
it has been suggested that the extrapolation is possible only because of the
mean-field approximation, and would not be possible in real systems due to a
singularity in the thermodynamic functions \cite{Fisher}. The presence of the
stable phase above some critical size in the metastable state is responsible
for the decay of metastable states and for the \emph{essential singularity} in
the free energy \cite{Fisher}. The singularity is absent in mean-field
theories or theories with long-range interactions. Accordingly, the existence
of SMS (no time-dependence) is commonly considered a mean-field consequence or
due to long-range interactions so that one should not seen SMS$^{\prime}$s in
real systems \cite{Penrose}. On the other hand, what one observes in
experiments are (time-dependent) metastable states. Since essential
singularities are almost impossibe to detect experimentally, it is not
surprising that the extrapolation is possible, at least from the
experimentalist's point of view.

In many cases, metastable states like supercooled liquids and glasses can
remain stable for a long period of times \cite{Kauzmann,Goldstein}. This
should be contrasted with metastability at high temperatures in the liquid-gas
transition that do not share this property. Thus, for supercooled liquids,
$\tau_{\text{s}}>>\tau_{\text{f}}.$\ This can be undestood by the high
viscosity observed in supercooled liquids, which slows down the growth of the
stable phase nuclei. There is another remarkable difference. Supercooled
liquids such as viscous liquids usually do not (but very well could, as was
seen recently \cite{GujRC}) exhibit spinodals, while supercooled vapor and
superheated liquid invariably do. Rather, viscous liquids undergo a glass
transition at low temperatures, about two-thirds of their melting temperature
$T_{\text{M}},$ provided the liquid is cooled in a way that crystallization
does \emph{not} intervene$.$ Here, the crystal phase (CR) represents the more
stable phase, and care must be exercised to forbid its nuclei to form while
cooling the viscous liquid. This makes the decay of the metastable state even
less probable, and strengthens the inequality $\tau_{\text{s}}>>\tau
_{\text{f}}.$ Thus, it is safe to treat viscous liquids as stationary
metastable states (SMS's), which can then be described by equilibrium
thermodynamics under the \emph{restriction} that the crystal phase is not
allowed. It is these SMS's that are of interest in this work. It is the hope
that the study of SMS will throw some light on the properties of observed
metastable states in the form of viscous fluids. In particular, the
extrapolated free energy below the melting temperature can be used to describe
supercooled liquids.

However, even if extrapolation is possible, one must still argue that the
thermodynamic functions describe the stationary limit of experimentally
observed metastable states. Under what condition(s) can one demonstrate this
association to be valid?%

\begin{figure}
[ptb]
\begin{center}
\includegraphics[
natheight=5.958600in,
natwidth=7.947600in,
height=3.007in,
width=4.0015in
]%
{I5DPGQ01.wmf}%
\caption{Schematic form of the generic entropy functions for various possible
states.}%
\end{center}
\end{figure}

While there is no rigorous theory of such SMS's at present, there are some
valuable approaches available in the literature. One such approach to describe
SMS\ is to use the PL formalism of Penrose and Lebowitz (PL) \cite{Penrose}
using \emph{restricted ensemble} method, which we modify and adapt for our
case below. The modification is the following. The decay of the metastable
states (to the stable state) in the PL formalism\ will be \emph{completely}
suppressed in order to make them stationary. Thus, nucleation of the stable
phase will not be allowed \ in our study. This is consistent with Maxwell's
idea \cite{Maxwell} that to observe metastable states, we must ensure that the
stable phase is not present. The properties of the SMS are what PL call the
static or reversible properties \cite{Penrose}.

An alternative scenario for extrapolation is by analytically continuing the
eigenvalues of the transfer matrix as presented in \cite{Newman}, which
attempts to accomplish the same as the restricted ensemble does but in a
somewhat direct fashion.

\subsection{Schematic Entropy Functions under Continuation}

In the PL approach, only certain microstates out of all are allowed, the
prescription of which is discussed in \cite{Penrose}. The restricted
microstates are used to define a restricted partition function, which is then
used to study metastable states. This is schematically shown in Fig. 1, where
the curve OHAB represents the entropy function $S_{\text{ord}}(E)$\ for the
ordered crystal state, while DH$^{^{\prime}}$AO$^{^{\prime}}$K represents the
entropy $S_{\text{dis}}(E)$ associated with the disordered liquid state. The
entropy as a function of $E$ must be thought of as the entropy in the
microcanonical ensemble \cite{Guj1}, which must be at its maximum in the
equilibrium state. Since a SMS is not an equilibrium state in the unrestricted
ensemble, its entropy at some $E$ \emph{cannot} exceed the entropy of the
corresponding equilibrium state at the same $E$. It is clear, therefore, that
at lower energies, the ordered state must have higher entropy, while at higher
energies the disordered state must have higher entropy. On the other hand, if
a time-dependent metastable state is prepared under the constraint that the
stable phase is not allowed, then the entropy function of such a metastable
state will be represented schematically by FG. The three free energies
corresponding to the above entropy functions are shown in the inset. A
consequence of the entropy maximization principle noted above is that the free
energy $F_{\text{dis}}(T)$ of SMS\ \emph{cannot} be lower than the free energy
$F_{\text{ord}}(T)$ of CR at the same temperature $T.$ This explains the form
of the free energy in the inset. The slope of the tangent line HH$^{^{\prime}%
}$ gives the inverse melting temperature, while the slope of the tangent line
OO$^{^{\prime}\text{ }}$gives the inverse temperature at which the free energy
DO$^{^{\prime}}$CK in the inset is equal to the free energy of the crystal
phase at absolute zero\ ($T=0$). The slope of the entropy at K is shown to be
finite, as opposed to the infinite slope at O. This point will be discussed
further below.

The question that naturally arises is whether the above extrapolation is
possible. It should also be noted that the extrapolation of the free energy
does not guarantee that metastable states associated with this extension exist
in the model. This will become clear in the following. Thus, the other
important issue is to understand the condition under which the extrapolation
will represent the stationary limit of the metastable states that might be
observed. To answer these questions, we borrow ideas from both approaches
mentioned above and develop an approach, which is then tested by considering a
1-dimensional lattice model. This model has only nearest-neighbor
interactions, and is solved exactly by the use of the transfer matrix. We find
that the extrapolation can be carried out without any ambiguity to describe
stationary metastable states (SMS) in this case. Thus, stationary
metastability can exist even in non-mean-field theories and without long-range
interactions, which is our main result. We further show that the extrapolation
yields a thermodynamically stable SMS free energy, at least mathematically
(see below for details), all the way down to absolute zero$.$ However, the
continuation cannot represent any metastable state at very low temperatures
when the entropy becomes negative, and must be stopped. At this point, the
continuation must be replaced by what is conventionally called an ideal glass;
see below. This situation should be contrasted with the termination of a
metastable state in a spinodal. The point where the entropy vanishes is not a spinodal.

\subsection{Fundamental Postulate}

We assume the \emph{existence} of SMS's, so that the partition function (PF)
formalism can be applied. The need for the assumption is easy to understand.
At present, our understanding of whether equilibrium (lowest free energy)
states can be demonstrated to exist mathematically even in simple models is
too limited. We should recall that the existence of equilibrium states is
taken for granted as a \emph{postulate} in statistical mechanics and
thermodynamics, where it is well known that it is extremely hard to prove
their existence.\ We quote Huang \cite{Huang}$:$ \textquotedblleft Statistical
mechanics, however, does not describe how a system approaches equilibrium, nor
does it determine whether a system can ever be found to be in equilibrium. It
merely states what the equilibrium situation is for a given
system.\textquotedblright\ Ruelle \cite{Ruelle} notes that equilibrium states
are defined \emph{operationally} by assuming that the state of an isolated
system tends to an equilibrium state as time tends to +$\infty.$ Whether a
real system actually approaches this state cannot be answered.

The problem becomes more complicated for SMS's like supercooled liquids in
which, at least at low temperatures, the relaxation becomes very sluggish and
it is highly likely that the appropriate relaxation time $\tau_{\text{f}}$
indeed tends to +$\infty.$ In other words, such an SMS may not even be
observed in a finite amount of time, even though associated time-dependent
metastable states can certainly be observed. Even in this case, the study of
the long-time limit of metastable states still has a predictive value, and can
be carried out using the statistical mechanical formalism.

\subsection{Reality Condition}

For the microstates to exist in Nature, it is evident that $W(E),$ the number
of microstates of energy $E,$ must satisfy the \emph{reality condition}%
$\ W(E)\geq1$ [so that the entropy $S(T)\geq0$] even in the restricted
ensemble. However, a state with \emph{negative entropy} can emerge under
extrapolations of the free energy. If it happens that the extrapolation
results in a negative $S(T)$ at low temperatures, this will indicate that the
extrapolation no longer represents real microstates, and the system could not
be found in those microstates in Nature.

There are two $\emph{independent}$ aspects of thermodynamics and statistical
mechanics. The first one is the requirement of \emph{stability} according to
which thermodynamic quantities like the heat capacity, the compressibility,
etc. must never be negative. The other aspect, independent of the stability
criteria, is the \emph{reality} condition that ensures that such states occur
in Nature \cite{Note3}. The mathematical extension of the free energy of the
disordered phase, while always satisfying the stability criteria everywhere
($T\geq0$), need not satisfy the reality condition, as our example will show below.

\section{Equilibrium Formulation}

\subsection{Canonical Partition Function}

We consider a system composed of $N$ particles confined in a given volume $V$
and at a given temperature $T$. The canonical PF is given by
\begin{equation}
Z_{N}(T)\equiv Tr\text{ }W_{N}(E)\exp(-\beta E), \label{PF}%
\end{equation}
where $Tr$ is over all possible values of the energy, $W_{N}(E)$ is the number
of microstates of energy $E$ \cite{Note2}, and $\beta\equiv1/T,$ $T$ being the
system temperature in the units of the Boltzmann constant $k_{\text{B}}.$ We
do not explicitly show the volume-dependence. We also introduce
the\ adimensional free energy (without the conventional minus sign)
\begin{equation}
\Omega_{N}(T)\equiv\ln Z_{N}. \label{FreeEs}%
\end{equation}
For microstates to exist in Nature, $W_{N}(E)\geq1$; hence the corresponding
entropy $S_{N}(E)\equiv\ln W_{N}(E)$\ $\geq0$. Whether this remain true for
the analytic continuation remains to be seen.

\subsection{Thermodynamic Limit}

The thermodynamic limit is obtained by taking $N\rightarrow\infty,$ and $V$
$\rightarrow\infty,$ keeping $v\equiv V/N$ fixed. The limit is taken by
considering the sequence formed by
\[
\omega_{N}(T)\equiv(1/N)\Omega_{N}(T),
\]
for different values of $N$ as $N\rightarrow\infty.$ The volume must be
changed according to $V=vN.$ For proper thermodynamics, the limit of the
sequences must exist, which we assume and denote it by $\omega(T).$ The
corresponding Helmholtz free energy is $f(T)=-T\omega(T).$

In the following, we will usually suppress the index $N$ on various
quantities, unless necessary.

\subsection{Conditions for Equilibrium and Negative Entropy}

We assume the existence of an equilibrium crystal, which has its energy
$\ E=E_{0}$ at $T=0.$\ It also has the lowest free energy at low temperatures.
Since $E_{0}$ is an allowed energy$,$ we must surely have $W(E_{0})\neq0.$
Assuming $TS(T)\rightarrow0$ as $T\rightarrow0$, which is always true
according to the Nernst's postulate, we recognize that $E_{0}$ represents not
only the Helmholtz free energy but also the energy of the perfect CR at $T=0$.
Since $W(E)$ is non-negative, $Z$ is a sum of positive terms. As a
consequence, the following two principles of equilibrium are always satisfied.

\subsubsection{Principles of Equilibrium}

\begin{itemize}
\item \emph{ Maximization Principle }The PF $Z$ must be \emph{maximized} in
the thermodynamic limit. The maximum value of $Z(T)$ corresponds to picking
out the maximum term $e^{S-\beta E}$ in (\ref{PF}). This maximum term
corresponds to $E=\overline{E}.$

\item \emph{Stability Principle }The heat capacity, which is given by the
fluctuations in the energy is non-negative.
\end{itemize}

It should be stressed that the non-negativity of the heat capacity and the
maximization principle only require the \emph{positivity} of $W(E)$ ($\geq0);$
$W(E)$ $\geq1$ is not required$.$ Thus, both principles remain valid even if
the entropy becomes negative \cite{Note3}. The above principles of equilibrium
and reality are two \emph{independent} aspects. This observation is going to
be useful when we discuss the metastable states below.

\subsubsection{Principle of Reality}

Conventional statistical mechanics for a system in the thermodynamic limit
describes equilibrium states in Nature, for which the above two principles of
equilibrium, along with the \emph{principle of reality} ($W(E)$ $\geq1,S(E)$
$\geq0)$ must be satisfied \cite{Note2}. All these conditions may not be met
by metastable states. (Metastability does not occur in finite systems.) What
we will see that it is the reality condition that can be violated by
metastable states.

\subsection{Order Parameter}

The presence of a melting transition at $T_{\text{M}}$\ (the inverse of the
slope of HH$^{^{\prime}}$ in Fig. 1) means that the \emph{disordered
}equilibrium liquid (EL) phase above $T_{\text{M}}$ and the ordered CR below
$T_{\text{M}}$\ correspond to different values of the order parameter $\rho,$
which is traditionally defined in such a way that $\rho=0$ represents the
disordered phase and $\rho\neq0$ the ordered phase CR$.$ (Our example below
will show explicitly how the microstates can be divided into the two disjoint
classes.) We denote the free energy per particle above $T_{\text{M}}$ by
$\omega_{\text{dis}}(T)$\ [ $f_{\text{dis}}(T)=-T\omega_{\text{dis}}%
(T)],$\ and below $T_{\text{M}}$ by $\omega_{\text{ord}}(T)$\ [ $f_{\text{ord}%
}(T)=-T\omega_{\text{ord}}(T)],$ from which we can calculate the entropies,
and energies per particle
\begin{equation}
s_{\alpha}(T)\equiv-(\partial f_{\alpha}/\partial T),\;e_{\alpha}%
(T)\equiv-(\partial\omega_{\alpha}/\partial\beta), \label{sealpha}%
\end{equation}
$\alpha=$dis, ord, respectively, corresponding to the two states. From
$s_{\alpha}(T)\;$and $e_{\alpha}(T),$ we can construct the functions
$s_{\alpha}(e)\equiv s_{\alpha}[e_{\alpha}(T)],$ where $e=E/N$ in the
thermodynamic limit$.$ The extensive entropy functions ($s_{\alpha}$
multiplied by $N)$ are shown schematically in Fig. 1.

\section{Stationary Metastable States and Restricted Ensemble}

\subsection{PL Scheme}

We briefly review the restricted ensemble formalism developed by Penrose and
Lebowitz \cite{Penrose}, and the required modification to suit our purpose.
Let $e_{\text{CR,M}},$ and $e_{\text{EL,M}}$ denote the energies of the
coexisting phases CR\ and EL at the melting temperatute $T_{\text{M}};$ see
points H and H$^{^{\prime}}$ in Fig.1$.$ It is clear that $s_{\text{dis}}(e)$
and $s_{\text{ord}}(e)$\ constructed above certainly exist for $e\geq
e_{\text{EL,M}},$\ and $e\leq e_{\text{CR,M}},$ respectively$.$ Over this
range, we do not need to introduce the resticted ensembles. To obtain
$s_{\alpha}(e)$ beyond their respective range noted above, however, we need to
introduce the restricted ensembles \cite{Penrose}.

We begin by considering the case of finite but very large $N.$ From
$s_{\alpha}(e),$ we can determine the number of microstates $W_{\text{dis}%
}(E)=\exp[Ns_{\text{dis}}(E/N)]\geq1$ consistent with $\rho=0$ at high
temperatures (or energies $E\geq E_{\text{EL,M}}=Ne_{\text{EL,M}}$), and the
number of microstates $W_{\text{ord}}(E)=\exp[Ns_{\text{ord}}(E/N)]\geq1$
consistent with $\rho\neq0$ at low temperatures (or energies $E\leq
E_{\text{CR,M}}=Ne_{\text{CR,M}}$)$.$ (The equalities and inequalities are
defined upto thermodynamically insignificant terms.) Let us focus on
$W_{\text{dis}}(E)$ for $E\geq E_{\text{EL,M}}$, \ which contains only those
microstates that are disordererd and correspond to $\rho=0.$ These microstates
may contain a small number of clusters or nuclei of stable phase (CR), but
their sizes are limited by the correlation length, which remains \emph{finite}
since we are dealing with a first-order transition. Let $\xi_{\text{dis}}$ (in
the units of some average inter-particle distance) denote the maximum value of
the correlation length in the disordered phase. We now follow PL, and select
all distinct microstates of energies $E<E_{\text{EL,M}},$ in which there are
no nuclei of the stable phase of sizes larger than $\xi_{\text{dis}},$ and the
number of smaller clusters is not too large, i.e. is thermodynamically
insignificant to ensure that these configurations also correspond to $\rho=0;$
the check of this will be discussed below$.$\ We denote the number of these
microstates also by $W_{\text{dis}}(E).$ We can similarly extend
$W_{\text{ord}}(E)$ to $E>E_{\text{CR,M}}.$\ Thus, we can construct the two
entropy functions $S_{\text{ord}}(E)\equiv\ln W_{\text{ord}}(E),$ and
$S_{\text{dis}}(E)\equiv\ln W_{\text{dis}}(E)$ that overlap, and are shown
schematically in Fig.1.

\subsection{Required Extension}

Let $\ E_{0\text{ }}$denote the lowest energy in the system, which represents
the energy of the ordered phase at $\ T=0.$ Thus, $W_{\text{ord}}(E)\geq1.$
While $W_{\text{ord}}(E)$ certainly exists for microstate energies starting
from $E=E_{0},$ there is no guarantee that $W_{\text{dis}}(E)$ also exists
near $E=E_{0}$. Most probably, $W_{\text{dis}}(E)$ does not continue all the
way down to $E=E_{0}$. If it did, the energy of the disordered phase at
absolute zero would be $E_{0}$ (we assume that $TS_{\text{dis}}\rightarrow0$
as $T\rightarrow0$)$,$ the same as that of CR. This would most certainly imply
that they would coexist at $T=0$, each having the same volume; recall that we
are considering a fixed volume ensemble$.$\ While there is no thermodynamic
argument against it, it does not seem to be the case normally. Usually, the
most stable state at $T=0$ is that of a crystal. Moreover, it is an
experimental fact \cite{Kauzmann} that all glasses have much higher energies
or enthalpies compared to their crystalline forms at low temperatures. Thus,
we assume that the lowest possible energy $E_{\text{K}},$ see Fig. 1,\ for the
disordered state is larger than $E_{0}.$ In other words, the microstate number
$W_{\text{dis}}(E)$ has the following property:
\begin{subequations}
\begin{equation}
W_{\text{dis}}(E)\geq1\ \text{ \ \ for \ \ }E\geq E_{\text{K}}. \label{Wdis}%
\end{equation}
If the slope in Fig. 1 at K is finite, then there is no sigularity in
$S_{\text{dis}}(E)$ at K, and we can extend it to lower energies. We assume
this extension is possible and define the extended entropy function for $E\geq
E_{0}.$ We denote this extended entropy function by $S_{\text{dis}}^{\ast
}(E),$ and introduce $W_{\text{dis}}^{\ast}(E)=\exp[S_{\text{dis}}^{\ast
}(E)].$ The function $S_{\text{dis}}^{\ast}(E)$\ is identical to
$S_{\text{dis}}(E)$\ over $E\geq E_{\text{K}}.$ It exists over the entire
range $E\geq E_{0},$ whereas $S_{\text{dis}}(E)$\ exists only over the range
$E\geq E_{\text{K}}.$ We can similarly extend $W_{\text{ord}}(E)$ to
$E=E_{\text{J}},$ where $E_{\text{J}}$ is either equal to $E_{\text{Max}}$,
the maximum allowed energy in the system, or the location of the singularity
in $S_{\text{ord}}(E)$ so that the latter cannot be extended beyond it. We
denote this extension similarly by $W_{\text{ord}}^{\ast}(E).$\ In the
following, we are mostly interested in the extension $S_{\text{dis}}^{\ast
}(E).$

\subsection{Restricted and Extended Restricted PF$^{\prime}$s}

Using $W_{\text{ord}}(E)$, $W_{\text{dis}}(E),$ and their extended version
$W_{\text{ord}}^{\ast}(E)$, $W_{\text{dis}}^{\ast}(E)$ we introduce the
following \emph{restricted ensemble }PF$^{\prime}$s \cite{Penrose}:
\end{subequations}
\begin{subequations}
\begin{align}
Z_{\alpha}(T)  &  \equiv TrW_{\alpha}(E)\exp(-\beta E),\text{\ }%
\label{PFalpha0}\\
Z_{\alpha}^{\ast}(T)  &  \equiv TrW_{\alpha}^{\ast}(E)\exp(-\beta E),
\label{PFalpha1}%
\end{align}
$\alpha=$dis, ord, and the corresponding free energies
\end{subequations}
\begin{equation}
\Omega_{\alpha}(T)\equiv\ln Z_{\alpha}(T),~\Omega_{\alpha}^{\ast}(T)\equiv\ln
Z_{\alpha}^{\ast}(T). \label{Falpha}%
\end{equation}
The free energy per particle $\Omega_{\alpha}(T)/N$ is expected to possess a
thermodynamic limit as $N\rightarrow\infty$, which we have already introduced
earlier as $\omega_{\alpha}(T).$ The corresponding limiting free energy per
particle $\Omega_{\alpha}^{\ast}(T)/N$ will be denoted by $\omega_{\alpha
}^{\ast}(T).$

\begin{itemize}
\item \emph{Remark} The following remark is important to understand the
relationship between the starred and unstarred PF$^{\prime}$s. Let us consider
the disordered PF$^{\prime}$s. For temperatures so that the average energies
$\overline{E}_{\text{dis}}^{\ast}(T)$ and $\overline{E}_{\text{dis}}(T)$ are
greater than $E_{\text{K}},$ both partition functions are determined by the
microstates of energies above $E_{\text{K}},$ where the starred and unstarred
$W_{\text{dis}}^{\prime}$s are identical. Hence, for $T\geq T_{\text{K}},$ the
two PF$^{\prime}$s $Z_{\text{dis}}^{\ast}(T)$ and $Z_{\text{dis}}(T)$ are the
same so that their free energies are the same. They differ only below
$T_{\text{K}};$ while $Z_{\text{dis}}^{\ast}(T)$ exists there, $Z_{\text{dis}%
}(T)$ does not. Similarly, for temperatures so that $\overline{E}_{\text{ord}%
}^{\ast}(T)$ and $\overline{E}_{\text{ord}}(T)$ less than $E_{\text{J}},$
$Z_{\text{ord}}^{\ast}(T)$ and $Z_{\text{ord}}(T)$ are the same. Thus,%
\begin{subequations}
\begin{align}
\omega_{\text{dis}}(T)  &  =\omega_{\text{dis}}^{\ast}(T),\;\;\;\;\;T\geq
T_{\text{K}},\label{FreeEEL}\\
\omega_{\text{ord}}(T)  &  =\omega_{\text{ord}}^{\ast}(T),\;\;\;\;\;T\leq
T_{\text{J}}. \label{FreeECR}%
\end{align}
Here, $T_{\text{J}}$ is the temperature where $\overline{E}_{\text{ord}}(T)$=
$E_{\text{J}}.$ The free energies $\omega_{\text{dis}}^{\ast}(T)$ is defined
for all temperatures $T\geq0.$\qquad
\end{subequations}
\end{itemize}

As long as $W_{\alpha}(E)>0,$ and $W_{\alpha}^{\ast}(E)>0,$ the restricted
PF$^{\prime}$s are sum of positive terms. Therefore, the corresponding free
energies satisfy the two equilibrium conditions noted above. Consequently,
even the restricted and extended restricted PF$^{\prime}$s will never give
rise to unstable states.

It is clear that the \emph{global maximization} of the PF requires that%
\begin{align*}
\omega(T)  &  =\omega_{\text{dis}}(T)=\omega_{\text{dis}}^{\ast}%
(T),\;\;\;\;\;T\geq T_{\text{M}},\\
\omega(T)  &  =\omega_{\text{ord}}(T)=\omega_{\text{ord}}^{\ast}%
(T),\;\;\;\;\;T\leq T_{\text{M}}.
\end{align*}
The switchover from $\omega_{\text{dis}}(T)$ to $\omega_{\text{ord}}(T)$ at
$T_{\text{M}}$ makes $\omega(T)$ singular, as expected, due to the transition.

We consider the case when there is only one phase transition, the first-order
melting transition, in the system. The following point is to be noted as
discussed by Penrose and Lebowitz \cite{Penrose}. The resticted PF$^{\prime}$s
defined in (\ref{PFalpha0}) and (\ref{PFalpha1}) require that we add an extra
energy term in the energy of the system, which takes the value 0 if the
microstate belongs to the set $\alpha$, and +$\infty,$ if it does not. This
meets the PL criterion for "static" metastable states. The other two criteria
that PL\ require relate to the decay of metastable states, and does not have
to be imposed here anymore. Thus, the problem of two incompatible requirements
discussed by Penrose and Lebowitz \cite{Penrose} no longer is an issue.

A prescription to describe metastability using the PF\ formalism can now be formulated.

\subsection{\textbf{Metastability Prescription}}

We \emph{abandon} the above global maximization principle, and use
$\omega_{\text{dis}}(T)$ to give the free energy of the metastable disordered
phase (supercooled liquid) below $T_{\text{M}}$ and\ $\omega_{\text{ord}}(T)$
to give the metastable (superheated crystal) state free energy above
$T_{\text{M}}.$ Similarly, $s_{\text{dis}}(T),$ $e_{\text{dis}}(T)$ and
$s_{\text{ord}}(T),$ $e_{\text{dis}}(T)$ give the entropy and energy per
particle for the supercooled liquid and superheated crystal, respectively.

There are two possibilities for the extrapolation of the free energy. As said
above, unstable states are not possible in the restricted ensemble. Thus,
either the free energy terminates in a spinodal at a non-zero but finite
temperature, or it extrapolates to $T=0$ through $\omega_{\text{dis}}^{\ast
}(T)$ for the supercooled liquid ($T\rightarrow\infty$ through $\omega
_{\text{ord}}^{\ast}(T)$ for the superheated crystal). In this work, we are
only interested in the supercooled liquid.

It is easy to calculate the order parameter $\rho$ for $T<T_{\text{M}}$ for
the "disordered phase" by using $Z_{\text{dis}}^{\ast}(T)$ to check if we have
properly identified the set of disordered microstates above. Since all
microstates in $W_{\text{dis}}(E)$ contain only nuclei of the stable CR\ phase
of finite sizes, the argument of Fisher \cite{Fisher} about the origin of an
essential singularity no longer works, which requires nuclei of all sizes,
including infinitely large sizes. Thus, it is clear that $\omega_{\text{dis}%
}^{\ast}(T)$ can be used to describe the saught extrapolation of the free
energy below the melting temperature. The phase represented by $\omega
_{\text{dis}}^{\ast}(T)$ below $T_{\text{M}}$ will still correspond to a
disordered state ($\rho=0$ ). This is our required description of SMS in the
form of SCL by the PF\ $Z_{\text{dis}}(T)$ below $T_{\text{M}}.$

From the above discussion, it appears highly likely that the singularity in
$\omega(T)$ does not necessarily imply a singularity in either of its two
pieces $\omega_{\text{dis}}(T)$ and $\omega_{\text{ord}}(T).$ Both$\ $of them
can exist on either side of $T_{\text{M}}.$ From the above argument, we
conclude that the extrapolation used to define $Z_{\alpha}^{\ast}(T)$ is not a
consequence of any approximation (mean-field or otherwise). Our example below
is intended to give a concrete demonstration.

The form of the entropy functions $S_{\alpha}(E)$ shown in Fig. 1 is also
supported by all known observations \cite{Kauzmann,Goldstein}, exact
calculations \cite{GujCorsi,GujRC,Derrida}, from the arguments given above and
the calculation to be presented below. We note that
\begin{subequations}
\begin{align}
S_{\text{ord}}^{\ast}(E) &  <S_{\text{dis}}(E),\;\;\;\;\;E>E_{\text{M}%
},\label{CompS}\\
S_{\text{dis}}^{\ast}(E) &  <S_{\text{ord}}(E),\;\;\;\;\;E<E_{\text{M}%
},\label{CompS1}%
\end{align}
where $E_{\text{M}}$\ is the energy at A where $S_{\text{ord}}%
(E)=S_{\text{dis}}(E);$ see Fig. 1 $.$ The SMS corresponding to the stationary
SCL is defined by the branch H$^{^{\prime}}$ACK and its extention to $E_{0},$
which is not shown. Similarly, superheated CR is defined by the branch HAB and
its extension to higher energies. We note that, as shown, the entropy
$S_{\text{dis}}$ of the metastable branch goes to zero at $T_{\text{K}}$%
$>$%
0 corresponding to the finite slope at K. This behavior will be supported by
the exact calculation in the next section.

%

\begin{figure}
[ptb]
\begin{center}
\includegraphics[
natheight=4.080200in,
natwidth=5.936900in,
height=2.476in,
width=3.5898in
]%
{I5FJ1W03.wmf}%
\caption{The bond and the entropy densities. The bond density is a monotonic
function of \textit{T}, so that the stability is not violated. The entropy
becomes negative at low temperatures, where metastable state must be replaced
by an ideal glass.}%
\end{center}
\end{figure}

\section{Exact 1-d Calculation}

The calculation presented here follow the transfer matrix eigenvalue approach
of Newman and Schulman \cite{Newman}. We now consider a one-dimensional axis
spin model, which contains $m$-component spins $\mathbf{S}_{i}$ located at
site $i$ of the one-dimensional lattice of $N$ sites, with periodic boundary
condition ($\mathbf{S}_{N+1}$ =$\mathbf{S}_{1}$). Each spin can point along or
against the axes (labeled $1\leq k\leq m$) of an $m$-dimensional spin space
and is of length $\sqrt{m}:$ $\mathbf{S=(}0,0,..,\pm\sqrt{m},0,..0)$. The
spins interact via a ferromagnetic nearest-neighbor interaction energy ($-J$),
with $K\equiv J/T>0.$ The energy of the interaction is given by
\end{subequations}
\[
E=-J\sum_{i=1...N}\mathbf{S}_{i}\cdot\mathbf{S}_{i+1}.
\]
The PF is given by
\begin{equation}
Z_{N}(K,m)\equiv\left(  \frac{1}{2m}\right)  ^{N}\sum_{\;}\exp(-\beta
E)=\left(  \frac{1}{2m}\right)  ^{N}Tr\;\widehat{\text{T}}^{N}, \label{AxisPF}%
\end{equation}
where the first sum is over the $(2m)^{N}$ spin states of the $N$ spins and
$\widehat{\text{T}}\equiv\exp(K\mathbf{S\cdot S}^{\prime})$ is the transfer
matrix between two neighboring spins. The transfer matrix has the eigenvalues
\begin{equation}
\lambda_{\text{dis}}=u+2(m-1),\lambda_{\text{ord}}=v,\lambda=u-2,
\label{Eigen}%
\end{equation}
that are 1-fold, $m$-fold, and ($m-1$)-fold, respectively \cite{GujAxis}. Here
we have introduced the following
\[
x\equiv\exp(Km),u\equiv x+1/x,v\equiv x-1/x.
\]

We follow de Gennes \cite{deGennes,Gujn0} and provide an alternative and very
useful interpretation of the above spin model in terms of a polymer system, in
which each polymer has multiple bonds and loops. The valence at each site in a
polymer must be even. (The presence of a magnetic field will allow odd
valencies, which we do not consider here.) The high-temperature expansion of
the PF, which is given by%
\begin{equation}
Z_{N}(K,m)=\sum K^{B}m^{L}, \label{PolyPF}%
\end{equation}
describes such a polymer system, with $K\geq0$, and $m$ denoting the activity
of a bond and the activity for a loop, respectively, and $B\;$and$\ L$
denoting the number of bonds and the number of loops, respectively
\cite{Gujn0}. The empty sites represent solvent particles. The number of
polymers and the number of bonds and loops in each polymer are not fixed and
vary according to thermodynamics. In addition, there is no interaction between
polymers, and between polymers and solvent particles, so that the polymer
system in (\ref{PolyPF}) is an \emph{athermal} solution. The temperature $T$
of the spin system does not represent the temperature in the polymer problem,
as is well known \cite{deGennes,Gujn0}. We will see below that small $x$
corresponds to high temperatures where the disordered phase is present, and
large $x$ corresponds to low temperatures where the ordered and possible SMS
phases are present. Thus, decreasing $T$ amounts to going towards the region
where the ordered and metastable disordered phases are present. Let $\omega$
denote the limiting value as $N\rightarrow\infty$\ of
\begin{equation}
\omega_{N}\equiv(1/N)\ln Z_{N}(K,m)+\ln(2m), \label{AxisFE}%
\end{equation}
where we have added an uninteresting constant to get rid of the prefactor in
(\ref{AxisPF}). This is done because the number of microstates appears within
the summation in the spin model PF in (\ref{AxisPF}). Thus, the inclusion of
the prefactor will make the microstate entropy negative. The prefactor is,
however, required for the polymer mapping.

The importance of the polymer mapping is that we can take\ $m\geq0$ to be a
real number, even though non-integer $m$ makes no sense for a physical spin.
Thus, for non-integer values of $m$, only the polymer system represents a
physical system. For $m=1,$ the axis model reduces to the Ising model, while
for $m\rightarrow0$, it reduces to the a model of linear chains with no loops
\cite{deGennes,Gujn0}. The eigenvalue $\lambda_{\text{dis}}$ is dominant at
high temperatures for all $m\geq0$\ and describes the disordered phase. Its
eigenvector is
\[
\left\langle \chi_{\text{dis}}\right\vert =\sum_{i}\left\langle i\right\vert
/\sqrt{2m},
\]
where $\left\langle 2k\right\vert $ ( or $\left\langle 2k+1\right\vert $)
denotes the single-spin state in which the spin points along the positive (or
negative) $k$-th spin-axis. It has the correct symmetry to give zero
magnetization ($\rho=0$). For $m\geq1,$ $\lambda_{\text{dis}}$ remains the
dominant\ eigenvalue at all temperatures $T\geq0$. For $0\leq$ $m<1,$ the
situation changes and $\lambda_{\text{ord}}$ becomes dominant\ at low
temperatures $T<T_{\text{c}},$ or [$x\geq x_{\text{c}}=1/(1-m)$] where
$T_{\text{c}}$\ is determined by the critical value $x_{\text{c}}\equiv
\exp(Jm/T_{\text{c}});$ there is a phase transition at $T_{\text{c}}$. The
corresponding eigenvectors are given by the combinations
\[
\left\langle \chi_{\text{ord}}^{(k+1)}\right\vert =[\left\langle 2k\right\vert
-\left\langle 2k+1\right\vert ]/\sqrt{2},\;
\]
$(k=0,2,..,m-1)$ which are orthogonal to $\left\langle \chi_{\text{dis}%
}\right\vert $, as can be easily checked$.$ These eigenvectors have the
symmetry to ensure $\rho\neq0$. The remaining eigenvalue $\lambda$ is
$(m-1)$-fold degenerate with eigenvectors
\[
\left\langle \chi^{(k+1)}\right\vert =[\left\langle 2k\right\vert
+\left\langle 2k+1\right\vert -(\left\langle 2k+2\right\vert -\left\langle
2k+3\right\vert )]/\sqrt{4},
\]
$(k=0,2,..,m-2.)$ For $m>0,$ this eigenvalue is never dominant. For
$m\rightarrow0,$ it becomes degenerate with $\lambda_{\text{dis}}.$ Since the
degeneracy plays no role in the thermodynamic limit, there is no need to
consider this eigenvalue separately for $m\geq0.$

We now consider the limit $N\rightarrow\infty.$ The adimensional free energy
per site, which represents the osmotic pressure \cite{GujRC,GujEqState}, of
the high-temperature equilibrium phase is $\omega_{\text{dis}}(T)\equiv
\ln(\lambda_{\text{dis}}).$ It can be continued all the way down to $T=0,$
even though the equilibrium osmotic pressure has a singularity at
$x_{\text{c}}.$ Similarly, $\omega_{\text{ord}}(T)\equiv\ln(\lambda
_{\text{ord}})$ related to the low-temperature equilibrium phase can be
continued all the way up to $T\rightarrow\infty.$ To calculate the entropy
density, we proceed as follows. The bond and loop densities are given by
\begin{equation}
\phi_{\text{B}}\equiv\partial\omega/\partial\ln K,\text{\ \ \ \ }%
\phi_{\text{L}}\equiv\partial\omega/\partial\ln m, \label{density}%
\end{equation}
which are needed to calculate the entropy per site of the polymer system
\[
s^{(\text{P)}}=\omega-\phi_{\text{B}}\ln K-\phi_{\text{L}}\ln m;
\]
the superscript is to indicate that it is the polymer system entropy, and is
different from the spin system entropy $s^{(\text{S)}}=\partial T\omega
/\partial T.$ If we define $\omega$ without the last term in (\ref{AxisFE}),
then $\phi_{\text{L}}$ and $s^{(\text{P)}}$\ must be replaced by
$(\phi_{\text{L}}-1)$ and $(s^{(\text{P)}}-\ln2),$ respectively. This will not
affect any of the conclusions below.

In the following, we will be only interested in the polymer entropy. The
proper stability requirements for the polymer system are
\begin{equation}
(\partial\phi_{\text{B}}/\partial\ln K)\geq0,(\partial\phi_{\text{L}}%
/\partial\ln m)\geq0, \label{PolyStbl}%
\end{equation}
as can easily be seen from (\ref{AxisPF}), and must be satisfied even for SMS.
They replace the positivity of the heat capacity of the spin system, which no
longer represents a physical spin system for $0\leq$ $m<1.$ It is easy to see
from the definition of $s_{\text{dis}}^{(\text{P)}}$ that $(\partial
s_{\text{dis}}^{(\text{P)}}/\partial T)_{m}$ need not be positive, even if the
conditions in (\ref{PolyStbl}) are satisfied.

Let us compute $\omega$\ as $K\rightarrow\infty$ ($T\rightarrow0)$ for the two
eigenvalues $\lambda_{\text{dis}}$ and $\lambda_{\text{ord}}$. From
(\ref{density}), it is easy to see that $\phi_{\text{B}}\rightarrow mK$ for
both states as $T\rightarrow0.$ Thus, using $\omega=s^{(\text{P)}}%
+\phi_{\text{B}}\ln K+\phi_{\text{L}}\ln m,$ we have
\begin{equation}
\omega_{\text{dis}}(T)/\omega_{\text{ord}}(T)\rightarrow1\;\text{\ as
\ }T\rightarrow0. \label{EqFE}%
\end{equation}
This means that if the eigenvalue $\lambda_{\text{dis}}$ is taken to represent
the metastable phase above $x_{\text{c}}$, its osmotic pressure must become
equal to that of the equilibrium phase (described by the eigenvalue
$\lambda_{\text{ord}}$) at absolute zero. This is in conformiity with Theorem
3 in \cite{GujProof}. We take $\omega_{\text{dis}}(T)$ to represent the
SMS\ osmotic pressure below $T_{\text{c}}.$ We have also checked that
$Ts_{\text{dis}}^{(\text{S)}}\rightarrow0,$ as $T\rightarrow0.$\ 

We will only discuss the disordered polymer phase below for $0\leq$ $m<1$. It
is easily checked that the above stability conditions in (\ref{PolyStbl}) are
always satisfied for $\lambda_{\text{dis}};$ see, for example, the behavior of
$\phi_{\text{B}}$ in Fig. 2, where we have taken $m=0.7,$ and $J=1$. Since the
high-temperature disordered phase represents a physical system, it cannot give
rise to a negative entropy $s_{\text{dis}}^{(\text{P)}}$ above $T_{\text{c}}$;
however, its metastable extension violates the reality principle as shown in
Fig. 2, where its entropy $s_{\text{dis}}^{(\text{P)}}$ becomes negative below
$T_{\text{K}}\cong0.266,$ which is lower than the transition temperature
$T_{\text{c}}.$

We now make an important observation. As $m$ decreases (below 1), both
$T_{\text{K}}$ and $T_{\text{c}}$ ($T_{\text{K}}<T_{\text{c}})$ move down
towards zero simultaneously. As $m\rightarrow0,$ the equilibrium ordered phase
corresponding to $\lambda_{\text{ord}}$\ disappear completely, and the
disordered phase corresponding to $\lambda_{\text{dis}}$ becomes the
equilibrium phase. There is no transition to any other state. Thus, there is
no metastability anymore. Consequently, there is \emph{no} ideal glass
transition since there is \emph{no} other state more ordered than this state
any more, as argued above. Thus, our exact calculation confirms our earlier
conclusion that the existence of an ordered state is crucial for the existence
of the entropy crisis. The existence of an ordered state sets the zero of the
temperature scale by its minimum energy $E_{0}$. This scale then sets the
temperature $T_{\text{K}}$ of the lowest SMS energy $E_{\text{K}}>E_{0}$ to be
positive. \ 

We also observe that there is no singularity in $\lambda_{\text{dis}}$ or
$\omega_{\text{dis}}(T)$ at $T_{\text{c}}$, even though there is a phase
transition there.\ Similarly, there is no singularity in $\lambda_{\text{ord}%
}$ or $\omega_{\text{ord}}(T)$ at $T_{\text{c}}.$\ Thus, the thermodynamic
singularity in the equilibrium free energy does not necessarily create a
singularity in $\omega_{\text{dis}}(T)$ or $\omega_{\text{ord}}(T)$ at
$T_{\text{c}},$ as was discussed earlier. The existence of a singularity or
spinodal at some other temperature is a different matter.

\section{Discussion \& Conclusions}

\subsection{SMS \& Exact Calculations}

The transition between SMS and the ideal glass is not brought about by any
thermodynamic singularity at $T_{\text{K}};$ rather, it is \emph{imposed} by
the reality requirement. The ideal glass state does \emph{not} explicitly
emerge as a new phase in the calculation since it is a disordered phase
defined by the order parameter $\rho=0$. In this sense, the transition to the
ideal glass is a very special kind of transition, which does not seem to
belong to the class of phase transitions in which various phases emerge in the calculation.

The exact calculation, which is not mean-field calculation in principle, in
the previous section demonstrates the existence of SMS. Thus, it demonstrates
that our hypothesis of SMS existence is not vacuous. It also shows that the
free energy can be extrapolated below the melting temperature by the use of
the restricted PF, and that there is no essential singularity, a signature of
a first-order transition \cite{Fisher}. The free energy remains stable all the
way down to absolute zero. However, the mere existence of the stable
extrapolated free energy all the way down to $\ T=0$ does not mean that it
represents the free energy of a realizable metastable state. This becomes
evident when we consider the entropy of the extrapolated free energy. This
entropy drops rapidly, and goes through zero at $T_{\text{K}},$ and becomes
negative as the temperature is reduced. A genuine entropy crisis appears in
the SMS below $T_{\text{K}}$. At $\ T_{\text{K}},~f_{\text{dis}}^{\ast
}(T_{\text{K}})=E_{\text{K}},$ which is higher than the CR free energy $E_{0}$
at $\ T=0.$ Below $\ T_{\text{K}},$ the extrapolated free energy cannot
represent any real metsatable state and must be replaced by another free
energy branch, which is constant:~$f_{\text{IG}}(T)=E_{\text{K}}$ for
$\ T<T_{\text{K}}.$ It is shown by the dotted horizontal straight line at K in
the inset.This branch represents the free energy of the ideal glass (IG),
which is the phase below $T_{\text{K}}.$ We need to invoke an ideal glass
transition at this temperature in the model. The energy of the ideal glass is
$E_{\text{K}}.$This means that the ideal glass has a higher energy than the
crystal at absolute zero, in conformity with the experiments.

It is interesting to note that $T_{\text{K}}\rightarrow0,$ as $m\rightarrow0$,
so that the ideal glass transition disappears. This is not surprising, as
$T_{\text{c}}\rightarrow0$. Thus, the there is no ordered state anymore.

\subsection{No Entropy Crisis in the Equilibrium State}

The entropy crisis occurs only in the metastable state, and not in the
equilibrium state, even though we have not shown this explicitly here. The
entropy of the latter vanishes at $E_{0}$ with an infinite slope, as shown in
Fig. 1. Thus, the lowest energy $E_{0}$ determines the lowest allowed
temperature $T=0$ in the system, which is consistent with the Nernst-Planck
postulate. However, it is possible that the equilibrium free energy becomes
horizontal, so that the entropy vanishes, over a non-zero temperature range
$(0,T_{\text{C}})$ where the system is frozen$.$ Such a situation happens, for
example, in the KDP model and gives rise to a singularity at $T_{\text{C}}.$
This should be contrasted with the existence of the ideal glass transition in
the supercooled state, where its appearance is not accompanied by any
singularity in the SMS free energy. Replacing the unphysical SMS free energy
below $T_{\text{K}}$ by a frozen state is done by hand; it does not emerge as
part of the calculation. Indeed, as our calculation has shown, the ideal glass
transition disappears as $m\rightarrow0.$ In this limit, $T_{\text{c}%
}\rightarrow0.$ Thus, the "ordered" state corresponding to $\lambda
_{\text{ord}}$ disappears, and the disordered phase remains the equilibrium
state all the way down to $T=0.$ Thus, it is safe to conclude that equilibrium
state in any system will \emph{never} show an entropy crisis (at a positive
temperature). If any exact calculation for the free energy or the entropy
predicts an entropy crisis at a positive temperature, this will necessarily
imply that there must exist another state, the equilibrium state, for which no
entropy crisis should exist.

This observation has been crucial in a recent investigation of a dimer model
\cite{FedorDimer} in which the disordered phase underwent a first-order
transition to an equilibrium ordered phase. The ordered phase then gave rise
to an entropy crisis at a lower temperature, which forced us to look for
another equilibrium state, which was eventually discovered above the
temperature where the entropy crisis was found, so that the crisis occurred in
a metastable state (this time emerging form an intermediate ordered state).
Similar situation occured in more complex systems containing particles of
different shapes and sizes \cite{CorsiThesis}.

In summary, we have shown that stationary metastable states can appear in
exact calculations also. They do not only occur in mean-field calculations.

We would like to thank Andrea Corsi and Fedor Semerianov for various useful
discussions, and help with the first figure (Andrea Corsi).

\bigskip


\begin{thebibliography}{99}                                                                                               %


\bibitem {Penrose}O. Penrose and J.L. Lebowitz in \textit{Fluctuation
Phenomena}, ed. E.W. Montroll and J.L. Lebowitz (North-Holland, 1979).

\bibitem {Bragg}W.L. Bragg and E.J. Williams, Proc. Roy. Soc. \textbf{A145},
699 (1934).

\bibitem {Derrida0}B. Derrida, J. L. Lebowitz, and E. R. Speer, Phys. Rev.
Lett. 87, 150601 (2001).

\bibitem {Fisher}M.E. Fisher, Physics \textbf{3}, 255 (1967). J.S. Langer,
Ann. Phys. N.Y. \textbf{41}, 108 (1967). O. Landford and D. Ruelle, Commun.
Math. Phys.\textbf{13}, 194 (1969).

\bibitem {Kauzmann}W. Kauzmann, Chem. Rev. \textbf{43}, 219-256 (1948).

\bibitem {Goldstein}\textit{The glass transition and the nature of the glassy
state}, M. Goldstein and R. Simha, eds. Ann. N. Y. Acad. Sci. 279 (1976).

\bibitem {GujRC}P. D. Gujrati, S. S. Rane and A. Corsi, Phys. Rev E
\textbf{67}, 052501(2003).

\bibitem {Maxwell}J.C. Maxwell, Scientific Papers, ed. W.D. Niven (Dover, N.Y.
1965); p.425.

\bibitem {Newman}C.M$.$Newman and L. Schulman, J. Math. Phys. \textbf{18}, 23 (1977).\ \ \ \ 

\bibitem {Guj1}P. D. Gujrati, Phys. Rev. E \textbf{51}, 957 (1995).

\bibitem {Huang}K. Huang, \textit{Statistical Mechanics}, (second edition),
John Wiley, p. 125.

\bibitem {Ruelle}D. Ruelle. \textit{Statistical Mechanics}:\ \textit{Rigorous
Results}, Benjamin, Reading, Massachusetts (1969), p.1.

\bibitem {Note3}The stability criteria such as a non-negative heat capacity
that immediately follow from the PF\ formulation are independent of the
non-negative entropy requirement. Thus, it is possible for the SMS to have a
negative entropy over some temperature range. This will only means that such
states are not observable in Nature.

\bibitem {Note2}It is well known that in classical statistical mechanics, the
entropy in continuum space can become negative. This is true of the ideal gas
at low temperatures. From the exact solution of the classical Tonks gas of
rods in one dimension, one also finds that the entropy becomes negative at
high coverage. Thus, either quantum mechanics is needed or a lattice structure
is needed to replace the continuum spapce to make entropy positive. We assume
here that this has been done and that the entropy for realizable microstates
cannot be negative.

\bibitem {GujCorsi}P. D. Gujrati and A. Corsi, Phys. Rev. Lett. \textbf{87},
025701 (2001); A. Corsi and P. D. Gujrati, Phys. Rev. E \textbf{68}, 031502
(2003), cond-mat/0308555.

\bibitem {Derrida}B. Derrida, Phys. Rev. B\textbf{\ 24}, 2613 (1981).

\bibitem {GujProof}P.D. Gujrati, cond-mat/0309143.

\bibitem {GujAxis}P. D. Gujrati, Phys. Rev. B \textbf{32}, 3319 (1985).

\bibitem {deGennes}P. G. de Gennes, Phys. Lett. \textbf{38 A}, 339 (1972).

\bibitem {Gujn0}P. D. Gujrati, Phys. Rev. A \textbf{38}, 5840 (1988).

\bibitem {GujEqState}P. D. Gujrati, J. Chem. Phys. \textbf{108}, 6952 (1998).

\bibitem {FedorDimer}F. Semerianov and P.D. Gujrati, cond-mat/0401047.
F.Semerianov, Ph.D. Dissertation, University of\ Akron (2004).

\bibitem {CorsiThesis}A. Corsi, Ph.D. Dissertation, University of\ Akron (2004).
\end{thebibliography}
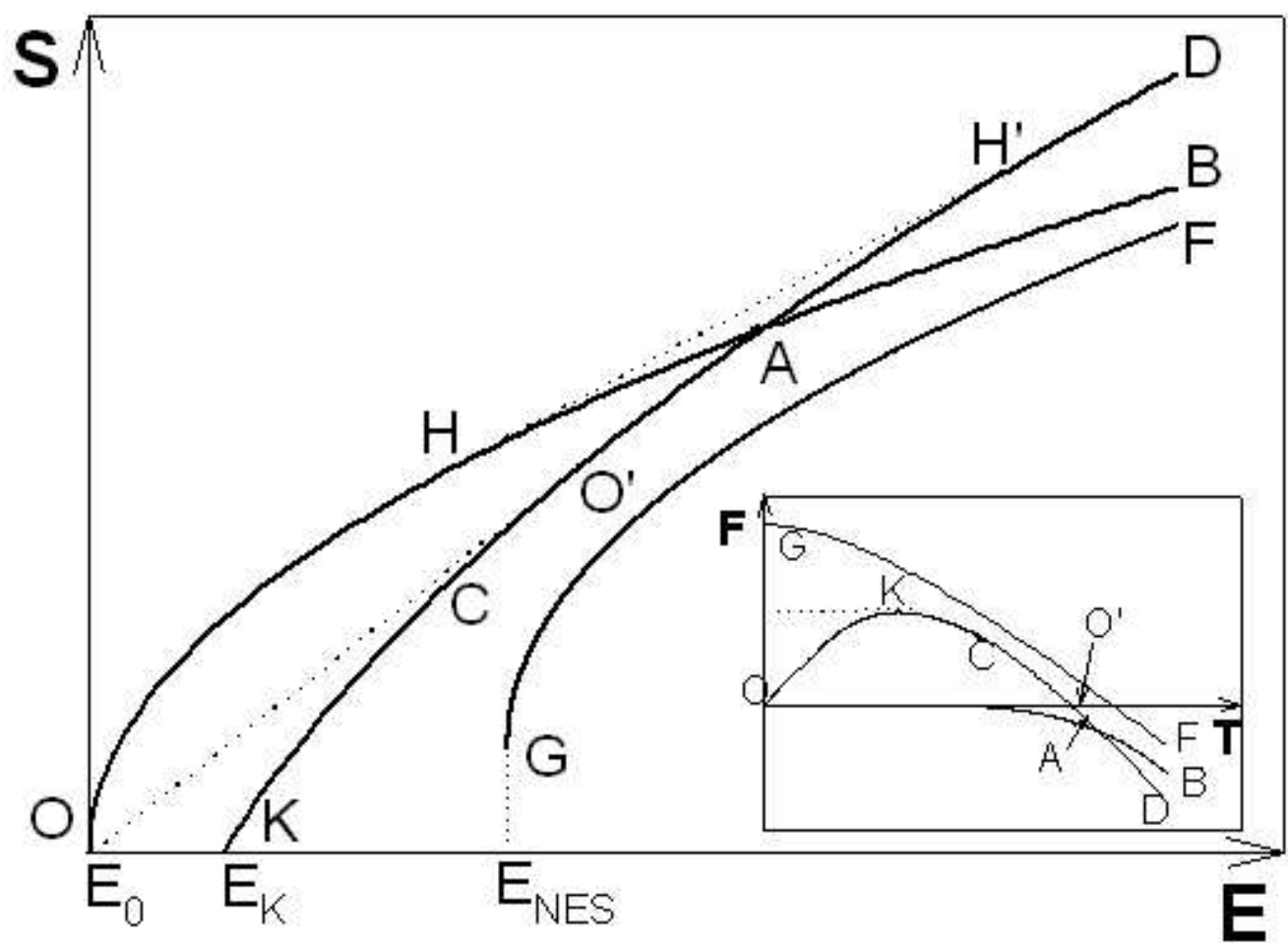
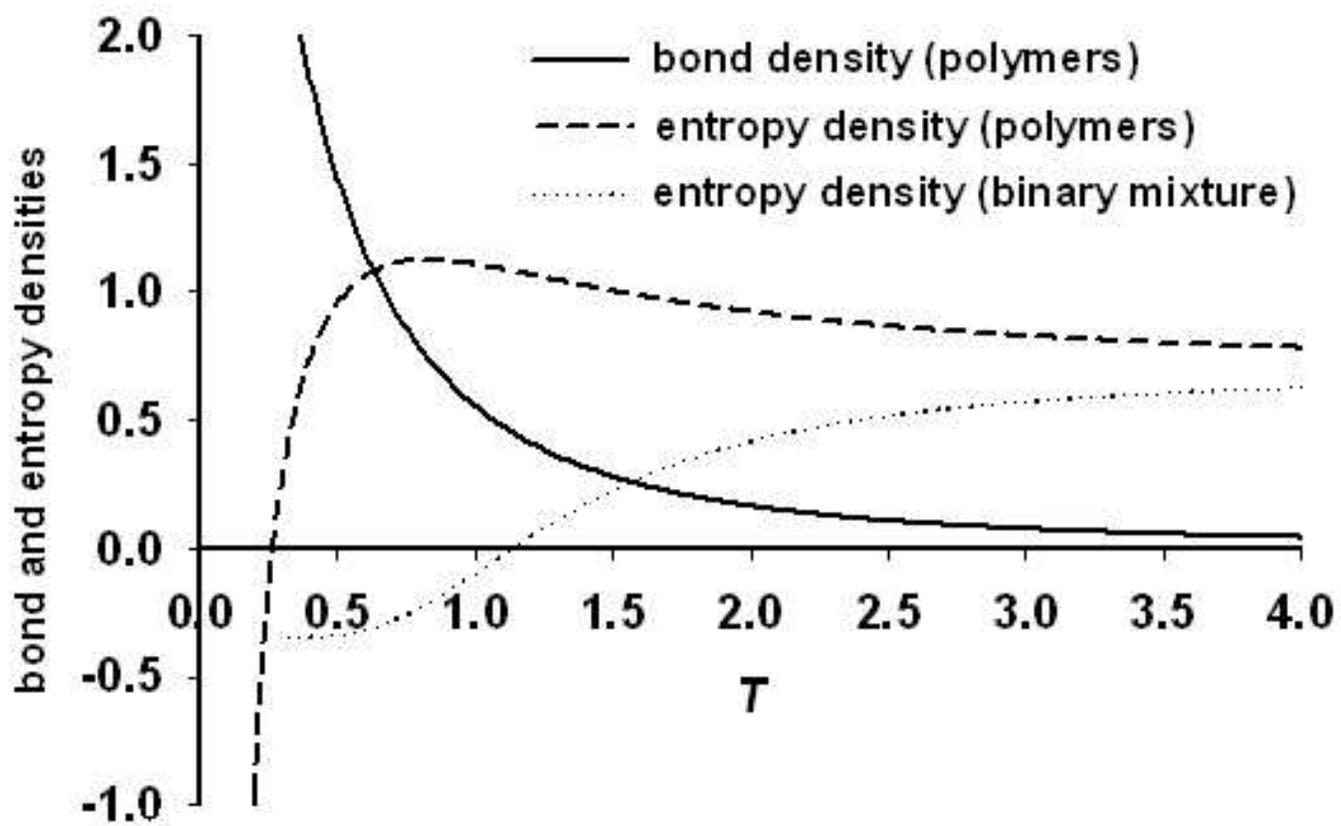
\end{document}